\documentclass{article}

\usepackage{arxiv}

\usepackage[utf8]{inputenc} % allow utf-8 input
\usepackage[T1]{fontenc}    % use 8-bit T1 fonts
\usepackage{hyperref}       % hyperlinks
\usepackage{url}            % simple URL typesetting
\usepackage{booktabs}       % professional-quality tables
\usepackage{amsfonts}       % blackboard math symbols
\usepackage{nicefrac}       % compact symbols for 1/2, etc.
\usepackage{microtype}      % microtypography
\usepackage{graphicx}
\usepackage{natbib}
\usepackage{doi}
\usepackage{subcaption}

\title{Decoding Imagined Handwriting from EEG}

\author{ 
Srinivas Ravishankar$^{1}$, Nora Zajzon$^{1}$, Virginia de Sa$^{1,2}$ \\
	Cognitive Science$^{1}$, Halicioglu Data Science Institute$^{2}$\\
	UC San Diego\\
    La Jolla, CA, USA \\ 
    \texttt{srravishankar@ucsd.edu}
}

\renewcommand{\headeright}{preprint}
\hypersetup{
pdftitle={A template for the arxiv style},
pdfsubject={q-bio.NC, q-bio.QM},
pdfauthor={Srinivas Ravishankar},
pdfkeywords={Electroencephalography (EEG), Brain-Computer Interfaces,  Handwriting, Decoding, Motor Imagery},
}

\begin{document}
\maketitle

\begin{abstract}
Patients with extreme forms of paralysis face challenges in communication, adversely impacting their quality of life. 
% Studies in decoding fine-grained dextrous movements such as handwriting from EEG are limited. 
Recent studies have reported higher-than-chance performance in decoding handwritten letters from EEG signals, potentially allowing these subjects to communicate. 
% However, our analysis shows that their performance can be primarily attributed to an ocular artifact caused by their experimental design. As such, it remains to be seen whether signals related to handwriting imagery can be reliably decoded from single-trial EEG. 
% In this work, we propose an improved experimental design that excludes the influence of confounding variables, and record simultaneous pen-tip trajectories and EEG activity while a healthy participant writes selected letters. We designed a decoder based on EEGNet and trained it on a subset of this data, with some letter instances held-out for evaluation. 
%First, we confirm that a prior study is affected by eye movement confounds. 
% Due to low Signal to Noise (SNR) ratio of EEG, ML methods might benefit from larger amounts of data. 
% We build upon prior work to investigate the limits of decoding performance that can be achieved. While prior work collected $\sim60$ trials per class for many subjects (breadth-wise exploration), we collect upto 600 trials per class on a limited set of subjects (depth-wise exploration). 
However, all prior works have attempted to decode handwriting from EEG during actual motion. Furthermore, they assume that precise movement-onset is known.
In this work, we focus on settings closer to real-world use where either movement onset is not known or movement does not occur at all, fully utilizing motor imagery. We show that several existing studies are affected by confounds that make them inapplicable to the imagined handwriting setting.
We also investigate how sample complexity affects handwriting decoding performance, guiding future data collection efforts. 
% We were able to achieve a maximum classification accuracy of \% between four letters. We were also able to decode the velocity time courses of held-out letter instances to form well-defined recognizable letters. 
%Finally, we also collect a smaller dataset with purely imagined handwriting and train a successful decoder on it. 
Our work shows that  (a) Sample complexity analysis in single-trial EEG reveals a noise ceiling, which can be alleviated by averaging over trials. (b) Knowledge of movement-onset is crucial to reported performance in prior works. (c) Fully imagined handwriting can be decoded from EEG with higher-than-chance performance.  Taken together, these results highlight both the unique challenges and avenues to pursue to build a practical EEG-based handwriting BCI.
\end{abstract}

% keywords can be removed

\section{Introduction}
\label{sec:introduction}

Patients with locked-in syndrome are unable to communicate through traditional means such as speech or writing, since their motor functions are severely impaired \citep{rousseau2013evaluation, laureys2005locked}. Brain Computer Interfaces (BCIs) that allow these subjects to communicate even in a limited fashion can significantly increase their quality of life \citep{milekovic2018stable, kubler2020history}. Many existing BCIs utilize motor imagery \citep{decety1996neurophysiological} as a control signal, which involves the subject attempting or imagining some familiar motor movement. The motor imagery paradigm is desirable as it allows for endogenous, asynchronous control, making it an ideal choice for everyday usage. Classical motor-imagery BCIs have used control signals such as left-hand/right-hand clenching, or tongue/hands/feet movement, etc \citep{mohamed2018comparison}. However, these types of control signals facilitate limited information transfer rates (ITR), and subjects must endure slow and cumbersome communication. 

Recently, imagined handwriting has produced significantly higher ITR than prior work, having been explored in the invasive setting \citep{willett2021high}. 
This previous study showed that the motor-imagery signals associated with handwriting can be reliably decoded using data from intracortical microelectrodes implanted in the precentral gyrus. In a single subject, pre-screened for high prior performance, this work demonstrated a real-time character error rate (CER) as low as 5.4\%, with a rate of communication approaching average smartphone typing speeds of the participant’s demographic.
% Interestingly, they found improved accuracy for letter classification than for center-out trajectories; the authors hypothesize that handwriting provides temporal as well as spatial differences in trajectories, allowing better separability of neural representations. 
% We hypothesize that this will also help with EEG-based classification of handwritten letters.
% Previous studies have sought to classify EEG activity evoked by motor imagery for cursor control and movement restoration through prosthetic limbs \cite{bright2016eeg}. 

\begin{figure}[h]
	\centering
	\begin{subfigure}{.3\textwidth}
		\centering
		\includegraphics[width=0.7\linewidth]{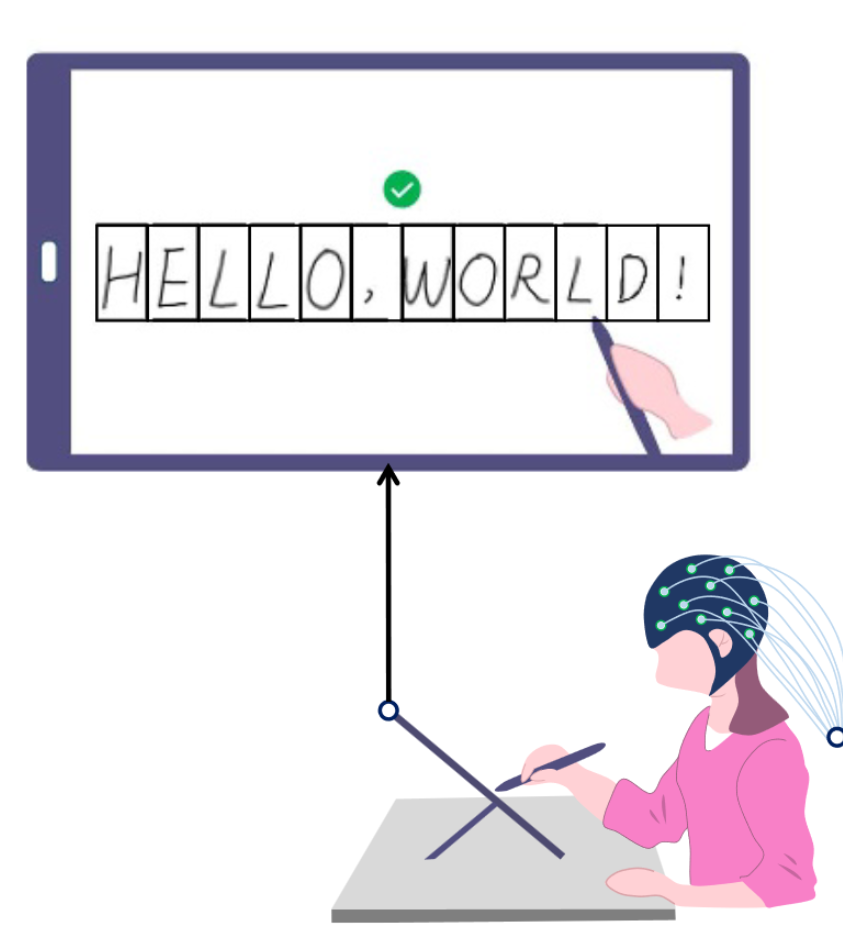}
		\caption{Prior work's exp. design}
		\label{fig:flawed_exp_design}
	\end{subfigure}%
	\begin{subfigure}{.7\textwidth}
		\centering
		\includegraphics[width=0.7\linewidth]{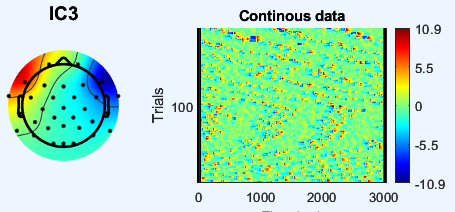}
		\caption{Eye IC}
		\label{fig:eye_IC}
	\end{subfigure}
	\caption{One of the 10 ICs used for prediction in prior work. The IC clearly represents the horizontal motion of the eyes tracking the letters in different boxes. This IC alone achieves 85.6\% accuracy.}
	\label{fig:prior_work}
\end{figure}

Inspired by this work, there have been a number of studies attempting to decode handwriting from Electroencephalography (EEG), a non-invasive brain modality. However, we find that many of these studies are affected by various confounds:
\citep{pei2021online} reported an accuracy of 94\% in a classification between the characters of “HELLO, WORLD!”. However, their experimental design involved writing the above characters in the same order every trial in specified boxes while looking at the tablet, shown in \ref{fig:flawed_exp_design}. Independent Components (ICs) representing eye-movements were not discarded from the EEG. 
% We conducted our own analysis on their data. 
The component shown in Fig \ref{fig:eye_IC} from ICA run on the EEG data likely represents horizontal eye movement, and was included in their predictive model. In our own analysis, excluding all other predictors, this artifact alone achieves 85.6\% accuracy in the classification task, while reported accuracy is 94\%. 
% Thus, it appears that the reported accuracy was obtained from the model decoding the subject's horizontal eye position, rather than motor signals associated with handwriting. 
 A later study \cite{kim2024towards} uses this dataset and achieves similarly high classification performance $\sim91\%$. It is likely that this work was similarly confounded by eye-movement artifacts owing to the experimental design.

Two works \citep{tripathi2024neuroair, jiang2025neural} collect EEG data for all 26 characters of the English alphabet during index-tracing and handwriting respectively. The reported performance was high, around $45\%$ and $31\%$ respectively on the entire letter set. 
In the first work, we find that using only electrodes \texttt{Fp1}, \texttt{Fp2}, \texttt{T8}, \texttt{TP10}, \texttt{P8}, average decoding performance is $\sim 24\%$, well above chance ($3.8\%$), indicating information from eye/shoulder motion is likely contributing to the high decoding performance. Furthermore, in their experimental design, the onset of the letter cue also acts as the writing cue. Thus, the letter is traced right after it appears on the screen; and trials are epoched around this onset. Recent works \cite{lan2023seeingbrainimagereconstruction, fei2025perceptogramreconstructingvisualpercepts} have shown that complex visual stimuli can be decoded from EEG. 
Thus, decoding performance may also be affected by visual decoding confounds. 
% We are unlikely to obtain this level of performance for an online BCI with a paralyzed subject, where no motion occurs and where the participant does not see a cued letter.
In the second work's experimental design, while the letter cue and the writing cue are separated, the subject received visual feedback on-screen as they traced the letter. A natural inclination while writing letters is to follow the pen tip with the eyes; potentially generating sensorimotor signals that are characteristic of each letter, and separate from the sensorimotor signals associated with handwriting. The authors show that gamma frequency and frontal electrodes are dominant in decoding performance. Recent work \cite{liu2025eeg2video} has also shown that complex video stimuli can be decoded from EEG, raising similar visual decoding possibilities as the previous work. 

In contrast, while \cite{crell2024handwritten} reports modest performance, $\sim 26.2\%$ over a 10 letter classification, their subjects trace the letter inside a box while fixating on the screen, and do not see any visual feedback related to the letter while they write. We emphasize that future work must follow these considerations to preclude confounds.
% Two related works \citep{tripathi2024neuroair, crell2024handwritten} both attempt to decode letters traced in the air using EEG. 
%We hypothesize that a more naturalistic handwriting motion using the highly innervated wrist and multiple fingers will elicit stronger signals for decoding, making it a more reliable motor imagery signal. Furthermore, our experimental paradigm involved collecting the precise pen-tip velocity during handwriting, alongside the EEG recording. This allows us to uncover new insights into the trained decoder at a granular level, as well as the motor encoding of handwriting signals, discussed further in Section \ref{sec:results}. 
% Both these works involve subjects tracing letters with their index-finger. 

However, a practical system for locked-in patients will require the ability to work with motor imagery, as in the intracortical study, rather than actual motion. While all prior work has dealt with real writing (on a tablet or in the air), our work investigates the unique challenges associated with decoding handwriting in settings where onset timing is unknown, or movement does not occur at all.

In addition, we conduct novel analyses in the setting employed by prior work, involving real motion and known onset timing. 
% We perform a sample complexity analysis that shows  and discover bottlenecks of decoding performance.
% Our work's experimental design is closest to this work, but we consider naturalistic handwriting on a tablet rather than index-tracing in the air.
% While this work involves finger tracing, we consider that naturalistic handwriting might be better encoded in the brain compared to index-finger writing.
% We thus collect pen-tip kinematics using a digital pen and tablet while simultaneous EEG is recorded as the subject is performing naturalistic handwriting. 
% We also collect data from both finger tracing and naturalistic handwriting from the same subject in a single session, to shed light on the neural representational differences between these, if any exist.
% With \cite{pei2021online} affected by confounds, and \cite{crell2024handwritten} reporting only modest decoding performance, a high performance EEG-based handwriting-BCI has not yet been demonstrated in literature. 
Invasive work \cite{willett2021high} has shown that scaling training data is a crucial step in achieving high performance, and we investigate if an EEG-based decoder similarly scales.
% While EEGNet is a compact architecture specializing in learning from small samples sizes, we investigate if decoding performance in EEG similarly scales with training data. 
To answer this question, we collect a large number of trials from a subject, to conduct sample-complexity analysis. 
% We find that in contrast to invasive works, the decoding model hits an early noise ceiling and saturates in classification performance. 
% Efforts to improve the paradigm might be necessary to overcome this potential limit of single-trial EEG performance.
In a complementary analysis, we also investigate if the SNR of single-trial of EEG acts as a bottleneck to decoding performance.

The contributions of our work are as follows:
\begin{itemize}
	% \item We collect EEG during Motor Execution (ME) and Motor Imagery (MI) handwriting tasks. Pen-tip velocity data is also collected during ME handwriting. We make this dataset publicly available.
    \item 
    % We analyze the relationship between sample complexity and EEG handwriting decoding performance. 
    We find that scaling training data, unlike invasive work, leads to diminishing gains in single-trial decoding. However, increasing SNR by averaging trials solely during evaluation improved performance in our best subject from $45\%$ to $78\%$ in a 4-way classification, a $73\%$ increase.
    \item We find that knowledge of onset timing is crucial to achieve reported decoding performance in literature; in a realistic setting where onset timing is unknown, performance drastically reduces, motivating future work aimed at addressing this gap.
    \item To the best of our knowledge, this work is the first to show that purely imagined handwriting can be decoded from single-trial EEG data with higher-than-chance performance.
\end{itemize}

% \noindent Taken together, this work illustrates the challenges and opportunities in building an EEG-based handwriting BCI closer to real-world use.

\section{Methods}
\label{sec:methods}

\subsection{Dataset}

\begin{figure}[h]
	\centering
	\begin{subfigure}{.3\textwidth}
		\centering
		\includegraphics[width=0.8\linewidth]{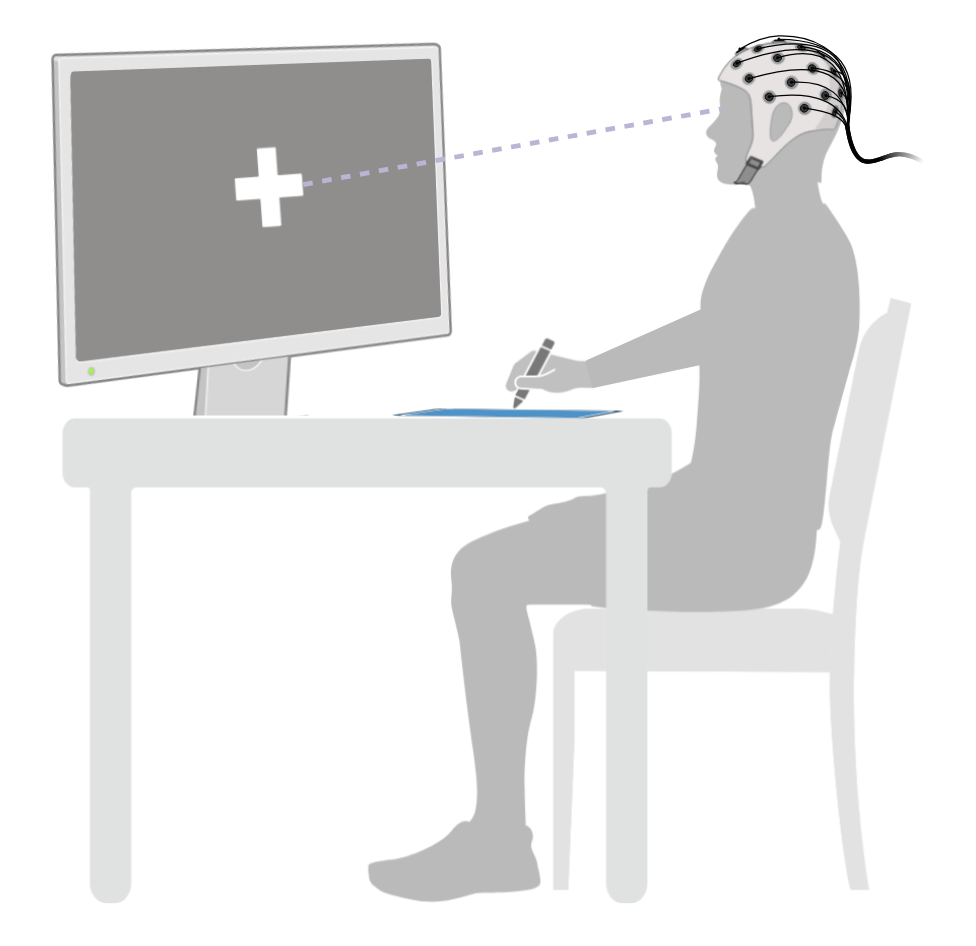}
		\caption{Experimental Setup}
		\label{fig:our_exp_setup}
	\end{subfigure}%
	\begin{subfigure}{.6\textwidth}
		\centering
		\includegraphics[width=\linewidth]{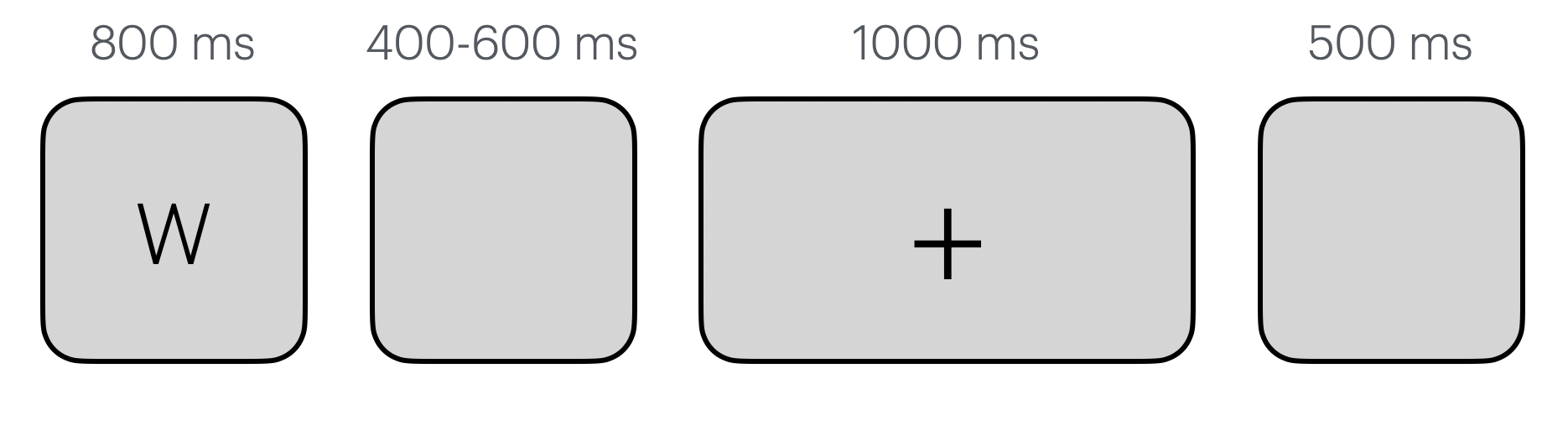}
		\caption{Block diagram}
		\label{fig:block_diagram}
	\end{subfigure}
	\caption{Experimental design for a single trial. The participant fixates on the monitor in front while writing on the tablet on the desk. One of the four letters is shown for 800 ms, followed by a blank screen for a randomly chosen period between 400-600 ms. Then a fixation cross appears on the screen for 1000 ms, during which the participant writes the letter on the tablet while looking at the cross. This is followed by a blank screen for 500 ms, after which the next trial begins.}
	\label{fig:experimental_design}
\end{figure}

Data was collected from four right-handed participants (2 male and 2 female) using 32 EEG channels. While our cohort size is quite limited compared to prior work \cite{crell2024handwritten}, we opted instead to collect more trials per subject and over multiple days. Our dataset can thus be used to analyze cross-session stability, with some sessions within a subject separated by more than 1 year. In place of the standard 10-20 montage, a custom montage was created to record more densely from the motor area of the brain, at the expense of fewer electrodes in the occipital and posterior regions of the head. The montage is provided in the supplementary. Cz was used as reference. 
% We proposed and employed a new  experimental design. 
Participants were instructed to fixate on a screen while they wrote on an android tablet placed on the desk in front of them, using a digital stylus as shown in Figure \ref{fig:our_exp_setup} in Motor Execution (ME) paradigm. In the Motor Imagery (MI) paradigm, they placed their hands on their lap instead. Thus, letters were independent of the location on the tablet, and participants' eyes did not track the letter as they were writing it. Four letters were chosen for this study, L, V, O and W, for two reasons: (a) These letters do not require the pen tip to leave the surface of the tablet, thus completely capturing the pen-tip motion corresponding to all muscle movements. (b) The letters allow interesting comparative analyses in future work, eg. L and V are rotations, V and W share the first part of their trajectories, etc. There are also a range of discontinuities exhibited by these letters, with W having three sharp discontinuities, L and V having one sharp discontinuity, and O having none. The block diagram of the experiment for a single trial is shown in Figure \ref{fig:block_diagram}.

Table \ref{tab:data_collection_sched} provides details about the sessions collected over different days, from different subjects, and under two possible timing paradigms. The number of trials were equally distributed among the four letters, presented in a pseudo-random order that ensured (a) Each 4 trials had all letters presented (b) No letter was presented twice in succession. 
The EEG data were collected at 1000 Hz, notch filtered at $60$ Hz, band-pass filtered between $0.3$ Hz and $70$ Hz, decomposed using ICA, and ICs representing muscle or eye artifacts were rejected by manual inspection. 
% band-pass filtered between $0.3$ Hz and $40$ Hz, 
Since this data had a maximum frequency of $40$ Hz, it was downsampled to 100Hz for efficient processing without loss of information. We also collected pen-tip trajectories using the tablet while the participant was writing the letters, at 200Hz. These trajectories were re-sampled to match the EEG data. Synchronization was done using photo-diodes, further details are given in supplementary.

\subsection{Decoder}

\begin{figure}[h]
	\centering
	\includegraphics[width=0.7\textwidth]{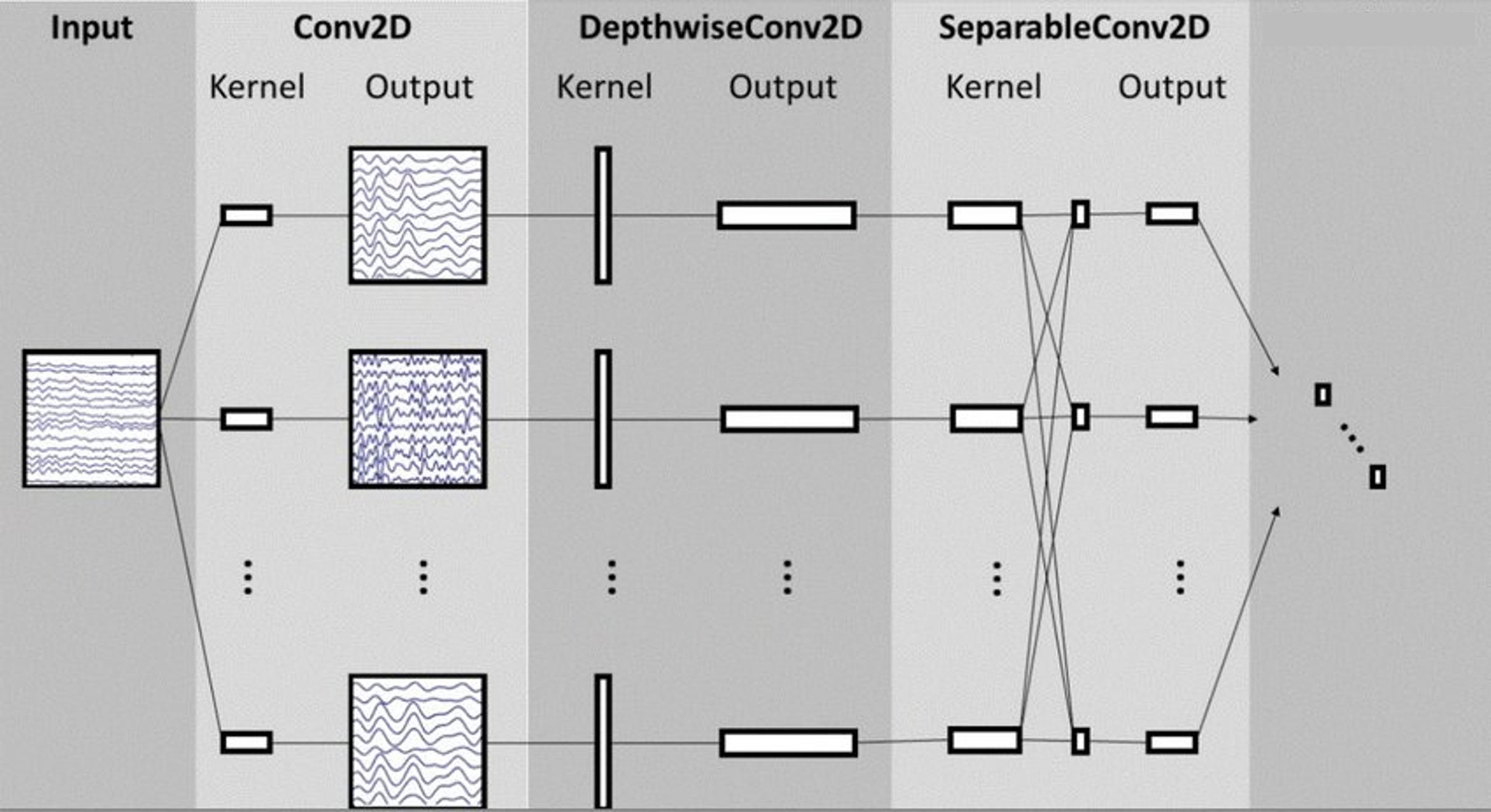}
	\caption{EEGNet architecture}
	\label{fig:eegnet}
\end{figure}

We use EEGNet-based models \citep{lawhern2018eegnet} as decoders, illustrated in Figure \ref{fig:eegnet}. 
% Following \cite{crell2024handwritten}, we coDirect classification: 
The model is trained to output a letter label corresponding to a snippet of EEG data
% , and (b) Kinematic decoding: The model is trained to decode part of or the whole pen-tip trajectory from the EEG data. The decoded trajectory is then fed to a simple trained RNN that classifies the trajectories into letters. 
% In either case, 
We quantify the final performance of the system by measuring its accuracy in classifying held-out letter instances as one of four letters (L, V, O and W). The chance performance is $25\%$ since the test data set is balanced.

\subsubsection{Direct Classification}
\label{sec:direct_classification}

We perform classification experiments under 3 settings: ME movement-centered, ME cue-centered, and MI cue-centered. In the first setting, we follow prior work and epoch trials around true movement onset. This is meant as a positive control, since prior work has shown that handwriting in this setting can be successfully decoded. 
In the second setting, we approach a realistic setting where real movement occurs, but precise movement onset is not known to the model. Trials are epoched around the cue to write letters, with timing to actual onset varying across trials.
Finally, the third setting is the most realistic setting possible in healthy subjects. Participants were asked to imagine writing the letters while they placed their palms on the lap. Since no actual motion occurred, all trials are cue-centered.

In all settings, 1000 ms of EEG data was used as input. In the movement-centered setting, trials were epoched [-200ms, 800ms] around movement onset. In cue-centered settings, trials were epoched [0ms, 1000ms] around the onset of the writing cue (fixation cross).
% For each letter, the EEG data were epoched from -200 ms to +800 ms after the pen touched the tablet. 
This length was chosen to be longer than the longest duration of letter writing in the data, which was $\sim$ 600 ms. 
We trained an EEGNet model with a four-way classification head, evaluated using 5-fold cross-validation. 
% We also investigate the setting where true movement onset is unknown,  to mimic a realistic setting with no actual movement. In this setting, the trials are epoched from 0 ms to +1000 ms after the fixation (writing) cue.

\subsubsection{Sample Complexity analysis}

This analysis seeks to answer the following question: How much data is required to train a functional BCI based on handwriting decoding from EEG? Given the low SNR of EEG data, it may be beneficial to collect more data than in comparable handwriting paradigms of invasive modalities. 
To answer these questions, we opted to collect a large amount of data per subject per class, rather than focusing on breadth of data (multiple letters with fewer trials per letter). In our largest dataset, from subject S1, we collect $\sim 2400$ trials, evenly split across 4 letters.  
% While we report k-fold cross validation accuracy in our previous , 
To facilitate a fair comparison across different training set sizes, we report performance on a fixed test set (final 160 trials of the dataset) rather than k-fold performance. The training super-set is composed of all prior trials. We investigate decoding performance as we vary the training set from 10\% to 100\% of its original size.

\section{Results}
\label{sec:results}

\subsection{Direct Classification}

\begin{figure}[h!]
	\centering
	\includegraphics[width=0.5\textwidth]{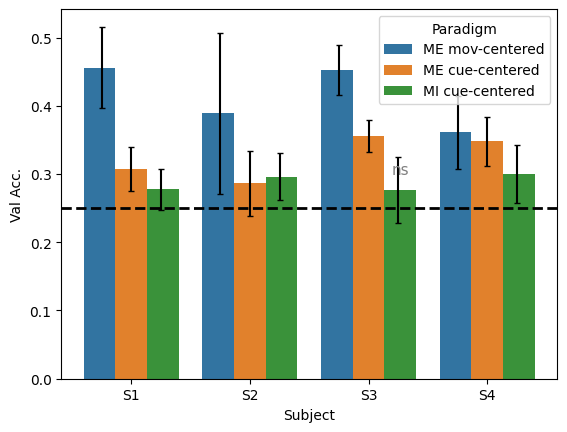}
	\caption{Subject-wise decoding performance on 3 different settings. Performance drops significantly when motor activity onset is unknown, even with actual motion; indicating a key issue to overcome for decoding imagined handwriting from EEG}
	\label{fig:results_consolidated}
\end{figure}

The performance of each subject in the three settings are shown in Figure \ref{fig:results_consolidated}. In line with prior work, all subjects achieve higher than chance accuracy in the ME movement-centered setting, and an overall average of $41.463 \pm 4.1$.
However, we report a significant drop in performance to $32.453 \pm 2.861$, when the trials are ME cue-centered, rather than ME movement-centered. Notably, data augmentation strategies such as random shifts of the input to make the model temporally invariant did not improve performance.
Finally, we show for the first time that imagined handwriting (MI) can be decoded with greater than chance performance, at $28.758 \pm 1.064$ (in all subjects except S3). While performance in this setting is also lower than the movement-centered ME setting, part of this drop is likely explained by what we observe in the fixation-centered ME setting.

\subsection{Sample Complexity}
\label{subsec:sample_complexity}

\begin{figure}[h!]
	\centering
	\includegraphics[width=0.5\textwidth]{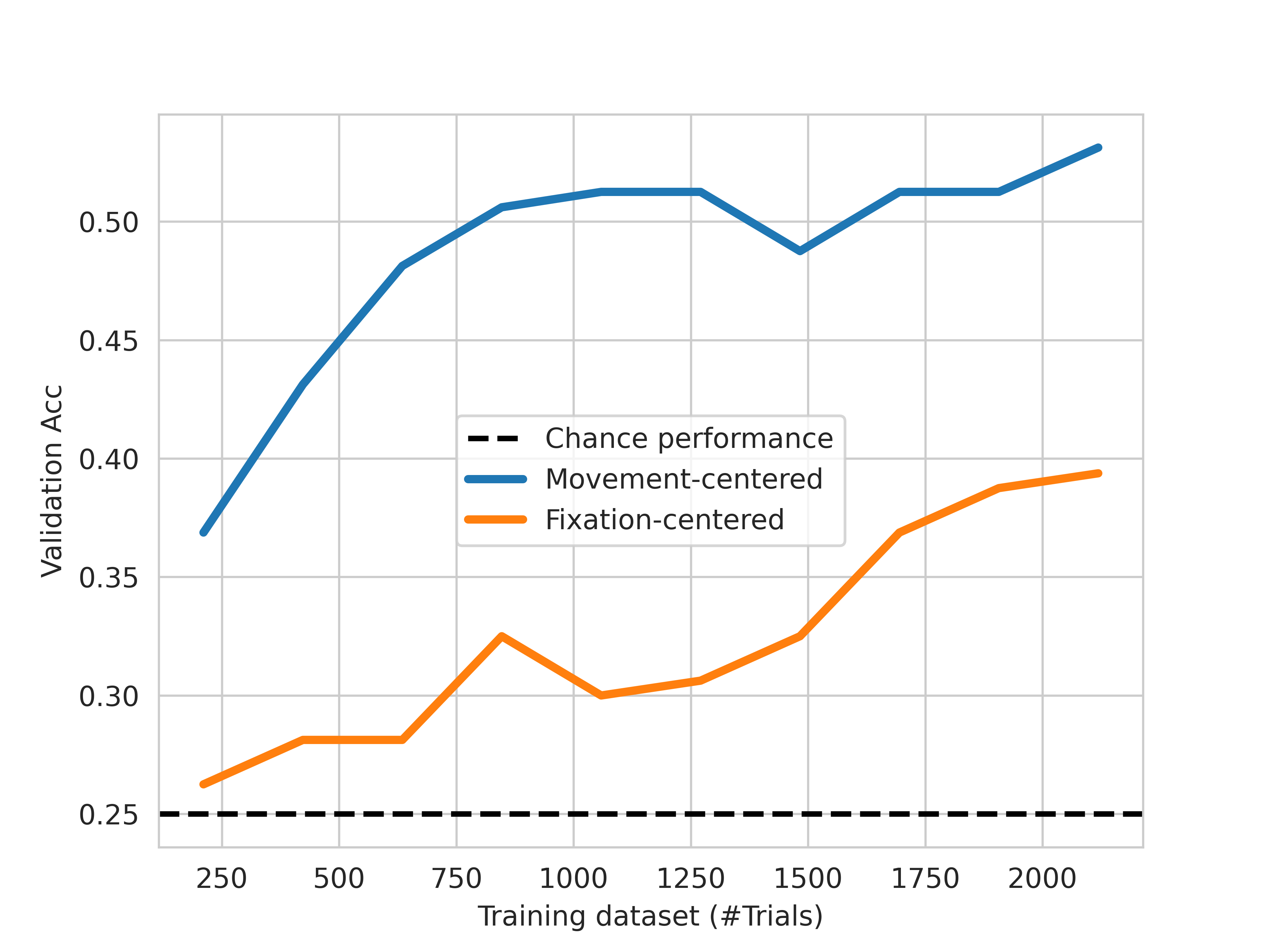}
	\caption{With knowledge of true movement onset (movement-centered), decoding performance exhibits diminishing gains as we scale up training data. Performance in the fixation-centered setting appears to follow this trend, albeit at a later stage.}
	\label{fig:sample_complexity}
\end{figure}

In this section, we investigate the effect of varying the size of the training dataset on the decoding performance. The performance on a held out set as we increase the training set size is shown in Figure \ref{fig:sample_complexity}. 
We observe that performance begins to saturate. 
In contrast to invasive works \cite{willett2021high}, scaling up training data provides diminishing gains on single-trial decoding performance. This result motivates the next section, where we investigate if the signal to noise ratio (SNR) of EEG is the bottleneck preventing high performance.

\subsection{Single-trial EEG SNR is the performance bottleneck}
\label{subsec:single_trial_SNR}

\begin{figure}[h!]
	\centering
	\includegraphics[width=0.5\textwidth]{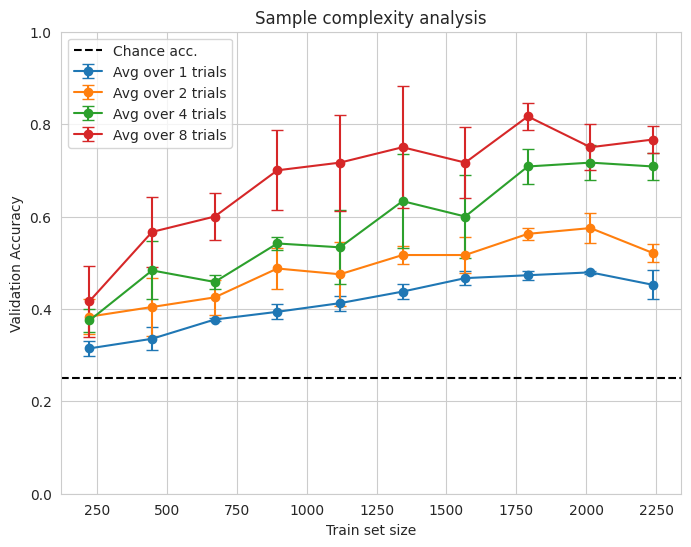}
	\caption{Boosting SNR during eval significantly improves performance even when training with single-trial EEG. Decoding performance on S1'
    s last 160 trials, when averaging over various number of trials of the same letter. Avg and std performance over 3 random seeds used for subsampling training sets.}
	\label{fig:SNR_boosted}
\end{figure}

In an invasive setting, single-trial data can be decoded with very high performance. However, the previous section showed that EEG-based handwriting decoding performance begins to saturate early even in a limited 4-letter classification. Here we investigate the effects of increasing SNR by averaging over multiple trials of the same letter (epoched around movement onset) during evaluation, while maintaining single-trial training as in the previous sections. These SNR-boosted sample complexity analyses when averaging over 2, 4, and 8 trials are contrasted with the previous single-trial performance in Figure \ref{fig:SNR_boosted}.

Averaging over input evaluation trials significantly improves performance, even when training is performed using single trials. In S1, this is a change from 45\% to as high as 78\% when averaging over 8 trials. 
% While these numbers are not directly comparable because they are effectively two different validation sets, 
Figure \ref{fig:SNR_boosted} clearly illustrates that the low SNR in single-trial EEG presents a noise ceiling for decoding performance, and is not likely to be alleviated by simply collecting more training data.

\section{Discussion and Limitations}
\label{sec:discussion}

Our results support recent work that motor information associated with handwriting can be decoded from EEG. Average single-trial decoding performance over subjects was $41.463\%$ in a 4-way letter classification. More importantly, we show that the precise timing of movement onset is crucial to achieving this performance. With real motion but unknown movement onset, performance dropped significantly. Crucially, this was not improved with data augmentation strategies that attempted to make the model temporally invariant. This may suggest that the letter-discriminative EEG activity is not itself very distinguishable from its temporally neighboring signals. 
% This motivates new approaches to capturing the onset of motor activity. 

Additionally, this work demonstrated that decoding handwriting from purely imagined handwriting is possible, but exhibits lower performance compared to movement-centered decoding with real movement. Notably however, the decoding performance is on par with the cue-centered ME trials, suggesting that solutions to the temporal invariance problem might result in significant performance gains.

On the other hand, this work shows that decoding performance even with perfect knowledge of movement onset saturates beyond $\sim 250$ trials per class. While this exact number may vary across subjects and letters chosen, it clearly demonstrates a noise ceiling in single-trial EEG that is far lower than decoding of handwriting from invasive recordings, where scaling up training data leads to very low error rates. While we showed that performance improves significantly by averaging multiple trials of the same letter during evaluation, it remains to be seen if such a solution leads to similar gains in a practical online setting, where movement onset is also unknown.

% Finally, given a latent space from a VAE trained on letter kinematics, we can generalize to unseen EEG patterns, provided those symbols' latent codes are spanned by other symbols with known EEG patterns. By identifying a subset of symbols spanning the latent space, this work demonstrates proof-of-concept in 0-shot handwriting decoding. For a larger symbol vocabulary in English or languages with many symbols (eg. Japanese with at least 70+ characters and 2000+ more symbols each denoting a complex concept), this could drastically reduce the amount of brain data collection necessary. 

% Non-invasive modalities can offer a safe and convenient alternative to surgical interventions.

\section{Conclusion and Future Work}
\label{sec:conclusion}

In this work, we showed that multiple prior studies on decoding handwriting signals from EEG overestimate performance of a realistic BCI that uses imagined handwriting. We built upon them to discover broader trends in decoding handwriting, both real and imagined. We showed that motor imagery can be decoded successfully in the pure motor imagery paradigm where no actual motion takes place.
Our results also guide future data collection efforts in terms of sample complexity requirements. 
% Future work will focus on extensive data collection to expand the classification to the full english letter set. A depth-wise data collection is recommended, with large number of trials collected from a few subjects. A naive transfer learning approach between subjects does not appear to improve performance, but more sophisticated approaches will be investigated.
% Motivated by the results in this work, we will pursue depth-wise data collection, acquiring a large number of trials per class for a single subject. Results on zero-shot encoding may allow for more efficient data collection. 

Future work will focus on methods to identify the onset of Motor Imagery, to bring performance on par to reported numbers in literature. Furthermore, paradigm changes are required to increase SNR of the signal before decoding. Future work will also involve comprehensive data collection to encompass the complete English alphabet, and study handwriting decoding in an online setting.
% However, we caution that repeated stimuli show dampened brain responses, and is likely to be non-trivial in an online setting.

% Finally, we demonstrate proof of concept in zero-shot decoding of letter kinematics from EEG, when no EEG data for a letter is seen. 
Taken together, our results highlights challenges and promising avenues associated with building a high-functioning EEG-based handwriting BCI.

\bibliographystyle{unsrtnat}
\bibliography{references}  %%% Uncomment this line and comment out the ``thebibliography'' section below to use the external .bib file (using bibtex) .

\newpage
%% The Appendices part is started with the command \appendix;
%% appendix sections are then done as normal sections
\appendix
\section{}
\label{app1}

\subsection{Montage}

\begin{figure}[h!]
\centering
\includegraphics[width=0.6\textwidth]{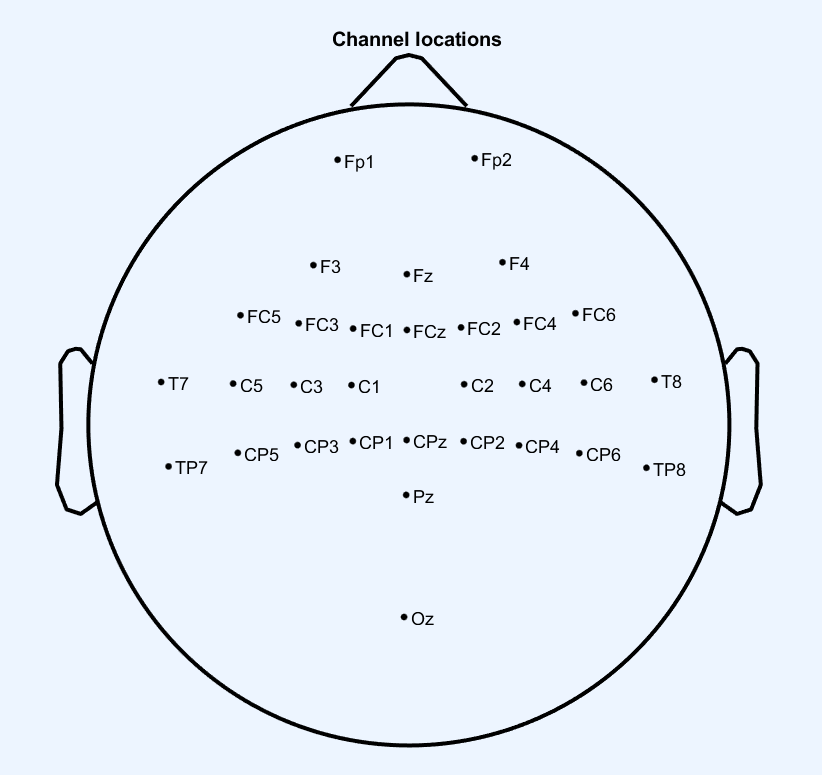}
\caption{A custom montage designed to record more densely from the midline area, over the motor cortex. Cz as reference.}
\label{fig:custom_montage}
\end{figure}

Data was collected from 32 EEG channels. Instead of the standard 10-20 montage, a custom montage was designed to record more densely from the mid-line area, as shown in Figure \ref{fig:custom_montage}. 
This was done to better capture signals from the motor cortex, at the expense of fewer electrodes in the occipital and posterior regions of the head. Cz was used as reference.

\subsection{Synchronization}

Three streams of data needed to be synchronized: The EEG stream from the amplifier, task cues from the recording computer, and the pen trajectories from the tablet. The LabStreamingLayer (LSL) networking ecosystem was used to send and receive the data between the devices. However, network latency can lead to streams being offset.

To ensure precise synchronization, we connected two photo-diodes to the amplifier, attaching one to a section of the tablet screen, and one to a section of the experimental monitor.  
The task program and the trajectory recording program were designed to flash the photo-diode region whenever a task cue was being shown, or a trajectory was drawn. These photodiode spikes were perfectly synchronized to the EEG since they fed into the same amplifier. The data/timestamps that were recorded with some latency by the LSL protocol were then aligned to the EEG stream using the closest photodiode spike in it.

We found network latencies and offsets of upto $80$ ms that were resolved using the photodiode-based synchronization.

\subsection{Data collection}

\begin{table}[h]
	\centering
	\begin{tabular}{ c|c|c } 
		\hline
		Subject & Session date & N Trials \\
		\hline
		\hline
		S1 & 23 Mar 2023 & 315  \\ 
		S1 & 26 Apr 2023 & 696  \\ 
		S1 & 4 Mar 2024 & 1268  \\
        S1 & 1 Jan 2025 & 1200  \\
        S1 MI & 1 Jan 2025 & 1240  \\
        S1 & 11 Jan 2025 & 120 \\
        S1 MI & 11 Jan 2025 & 400  \\
        \hline 
		S2 & 25 Feb 2024 & 339  \\ 
		S2 & 27 Feb 2024 & 336  \\ 
		S2 & 20 Oct 2024 & 778  \\ 
        S2 MI & 29 Dec 2024 & 1198 \\
        S2 & 20 Jan 2025 & 319 \\
		\hline
		S3 & 3 Nov 2024 & 474 \\ 
		S3 MI & 3 Nov 2024 & 300 \\ 
        \hline
		S4 & 7 Nov 2024 & 400  \\ 
		S4 MI & 7 Nov 2024 & 400 \\ 
		
		\hline
	\end{tabular}
	\caption{Details of data collection schedule}
	\label{tab:data_collection_sched}
\end{table}

\end{document}